\documentclass[onecolumn]{revtex4-1}
\usepackage[utf8]{inputenc}
\usepackage{amsmath,natbib}
\usepackage{amsfonts}
\usepackage{amssymb}
\usepackage{graphicx}
\everymath{\displaystyle}
\usepackage{multirow}
\usepackage{units}
\usepackage[squaren]{SIunits}
\usepackage{epstopdf}
\everymath{\displaystyle}
\newcommand{\bigO}[1]{\ensuremath{\mathop{}\mathopen{}O\mathopen{}\left(#1\right)}}

\newcommand*\mean[1]{\overline{#1}}
\newcommand{\myvec}[1]{\mathbf{#1}} %{\vec{#1}}
\newcommand{\mycit}[1]{\cite{#1}} % \citealt \citet \citenum
\newcommand{\minitab}[2][l]{\begin{tabular}{#1}#2\end{tabular}}
\begin{document}
\title{Optimal Energy Harvesting from Vortex-Induced Vibrations of Cables}
\author{Guillaume O. Antoine} 
\affiliation{LadHyX -- D\'epartement de M\'ecanique, Ecole Polytechnique -- CNRS, 91128 Palaiseau, France.}
\author{Emmanuel de Langre}
%\email{delangre@ladhyx.polytechnique.fr}
\affiliation{LadHyX -- D\'epartement de M\'ecanique, Ecole Polytechnique -- CNRS, 91128 Palaiseau, France.}
\author{S\'ebastien Michelin}
\email{sebastien.michelin@ladhyx.polytechnique.fr}
\affiliation{LadHyX -- D\'epartement de M\'ecanique, Ecole Polytechnique -- CNRS, 91128 Palaiseau, France.}
\date{\today}
 
\begin{abstract}
Vortex-induced vibrations (VIV) of flexible cables are an example of flow-induced vibrations that can act as energy harvesting systems by converting energy associated with the spontaneous cable motion into electricity. This work investigates the optimal positioning of the harvesting devices along the cable, using numerical simulations with  a wake oscillator model to describe the unsteady flow forcing. Using classical gradient-based optimization, the optimal harvesting strategy is determined for the generic configuration of a flexible cable fixed at both ends, including the effect of flow forces and gravity on the cable's geometry. The optimal strategy is found to consist systematically in a concentration of the harvesting devices at one of the cable's ends, relying on deformation waves along the cable to carry the energy toward this harvesting site. Furthermore, we show that the performance of systems based on VIV of flexible cables is significantly  more robust to flow velocity  variations, in comparison with a rigid cylinder device. 
This results from two passive control mechanisms inherent to the cable geometry : (i) the adaptability to the flow velocity of the fundamental frequencies of cables through the  flow-induced tension and (ii) the selection of successive vibration modes by the flow velocity for cables with gravity-induced tension.

\end{abstract}
\maketitle

\section{Introduction}

The field of renewable energies is gaining interest due to the limited availability and the environmental impact of fossil fuels. Flow-induced vibrations, i.e. the  motion of a solid structure resulting from the destabilizing effect of forces applied by the surrounding flow, have recently received an increased attention as a potential alternative to classical wind- and water-turbines to convert a fraction of the fluid's kinetic energy into electricity~\mycit{bernitsas2008vivace}. 

From an energy point of view,  flow-induced vibrations  transfer a fraction of the kinetic energy of the incoming flow to the solid structure that is set into motion. Vibrations can then be used to power an electric generator and effectively convert some of the solid's mechanical energy into electrical form. Recently, many classical examples of flow-induced vibrations have been revisited as potential energy harvesting mechanisms, including galloping~\mycit{barrero2010gallopingharvest, jung2011expvalidharvestfromgallop, abdelkefi2013piezoharvestfromgalloptriangbeam, vicente2014optimalEMharvestfromgallop, dai2015gallopelectromag}, coupled-mode flutter of airfoil profiles~\mycit{xiao2014}, flutter of flexible cylinders or membranes in axial flow~\mycit{singh2012biarticulatedflag, singh2012continuousflag, doare2011linearpiezoflag, michelin2013nonlinearpiezoflag}, flapping in unsteady wakes~\mycit{allen2001piezoshellincylinderwake, taylor2001piezobeamincylinderwake, akaydin2010piezobeamincylinderwake, wang2012diaphragmKarmanvortex} and vortex-induced vibrations (VIV) of rigid and flexible structures~\mycit{bernitsas2008vivace, meliga2011harvestVIV, hobbs2012treeVIV, barrero2012harvestVIV, akaydin2012vivpiezoharvesting_exp, mehmood2013piezoVIV, grouthier2014articleJFS, dai2014vivpiezoharvesting_theory}. The present work focuses on the possibility to harvest energy from VIV of long flexible cables. 

\begin{figure}[h!]
\begin{center}
\includegraphics[width=.8\textwidth]{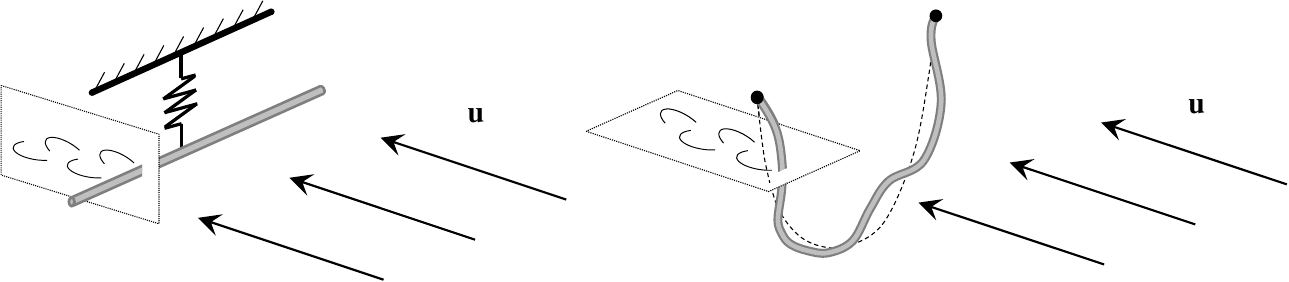}
\caption{Vortex-induced vibrations of a rigid cylinder (left) and a flexible cable (right) in a steady and uniform cross flow.}
\label{fig:intro_pic_VIV_rigid_cable}
\end{center}
\end{figure}

VIV result from the coupling of a bluff body to its unsteady vortex wake. The steady flow around a fixed bluff body at high Reynolds number ($\mbox{Re}\gg 1$ or inertial flows) is characterized by an unsteady shedding of vortex structures.  The wake structure is then  complex and differs from a classical Von-Karman vortex street observed at lower $\mbox{Re}$, but its frequency spectrum is still dominated by a fundamental frequency known as the Strouhal frequency $f$, proportional to the flow velocity $U$ as $f=S_T\,U/D$ with $S_T$ the Strouhal frequency and $D$ the diameter (e.g. \mycit{sarpkaya2004reviewVIV,williamson2004reviewVIV}). This results in an unsteady lift force on the bluff body. For a flexible or flexibly-mounted structure (see Fig.~\ref{fig:intro_pic_VIV_rigid_cable}), this unsteady force will force the solid body into so-called vortex-induced vibrations, and this unsteady motion of the body will also introduce a feedback coupling on the vortex shedding and unsteady lift. VIV of a rigid cylinder mounted on an elastic foundation have been widely studied, both experimentally and numerically (see~\mycit{sarpkaya2004reviewVIV,williamson2004reviewVIV,williamson2008briefreviewVIV,paidoussis2010}. A distinctive feature of VIV is the lock-in mechanism: when the fundamental frequency of the solid's vibrations is close to the Strouhal frequency, the coupling between the vibrations and the vortex shedding synchronizes both dynamics  over an extended range of flow velocity. This results in self-sustained, self-limited large amplitude oscillations of the solid, typically of the order of one cross-flow diameter.

This lock-in phenomenon is particularly interesting for energy harvesting purposes since large amplitudes imply that the oscillating structure has a large amount of kinetic energy that can potentially be harnessed to produce electricity. The VIVACE system \mycit{bernitsas2008vivace} relies on that phenomenon.

However, VIV of elastically-mounted rigid cylinders also have intrinsic limitations. Lock-in occurs only when the Strouhal frequency and the eigenfrequency of the structure are sufficiently close, e.g. \mycit{williamson2004reviewVIV}. If this condition is not met (``lock-out''), the rigid cylinder is not properly excited by the flow and its oscillations have negligible amplitude, which results in inefficient energy extraction. Geophysical flows (e.g. oceanic, tidal or river currents) are characterized by an important variability in the flow velocity magnitude, and systems relying on VIV of rigid cylinders can only produce energy when the flow velocity remains close to the velocity for which they are designed. 

A method to circumvent this issue is to use flexible structures (cables,  or beams) rather than rigid cylinders, as sketched in Fig.~\ref{fig:intro_pic_VIV_rigid_cable}: cables have multiple deformation modes which have their own eigenfrequencies and different modes can be locked-in for different values of the flow velocity (and Strouhal frequency). The selective excitation of a specific deformation mode of a flexible structure when the eigenfrequency of the latter is close to the Strouhal frequency was demonstrated experimentally~\mycit{king1995modeslockin,chaplin2005modeslockin} and numerically~\mycit{violette2010articleJFS}. Rather than lock-out, variations of the flow velocity induce a transition towards lock-in of higher or lower deformation modes.  This advantage of flexible structures was recently demonstrated for a system consisting of a hanging cable in a cross flow, attached to a local energy harvester at its upper end~\mycit{grouthier2014articleJFS}. The experimental and numerical results obtained for that system demonstrated its increased robustness to variations in the flow velocity in comparison with an elastically supported  rigid system.

 A challenge and open question associated with harvesting energy from VIV of flexible structures is to determine the optimal harvesting strategy, i.e. where and how the mechanical energy of the solid structure should be converted into electricity. With a rigid cylinder, the motion is a pure translation with a single degree of freedom, and determining the optimal strategy is relatively simple. For flexible structures, the periodic motion is not uniform throughout the structure, which results in a much larger configuration space for the harvesting distribution. Optimization of such a system is a particular challenge. Moreover, the geometry of the structure and how it is held in the flow are additional design variables. The present work addresses these questions and investigates the optimal route to efficient energy harvesting using VIV of flexible cables.

The paper is organized as follows. Section~\ref{sec:model} presents the mathematical and numerical models used here to analyse the VIV dynamics and resulting energy harvesting. In Section~\ref{sec:opti_harvest_strategy_3d}, the distribution of harvesting devices on a flexible cable is optimized, and the performance of such system is analysed and compared to the reference rigid system in Section~\ref{sec:discuss}. Section~\ref{sec:conclusion} finally summarizes the main conclusions of our work, stressing out both its fundamental and engineering implications.

\section{Model for the VIV of Hanging Cables \label{sec:model_static_VIVs}}
\label{sec:model}

\subsection{Problem geometry}
\begin{figure}
\begin{center}
\includegraphics[scale=1.00]{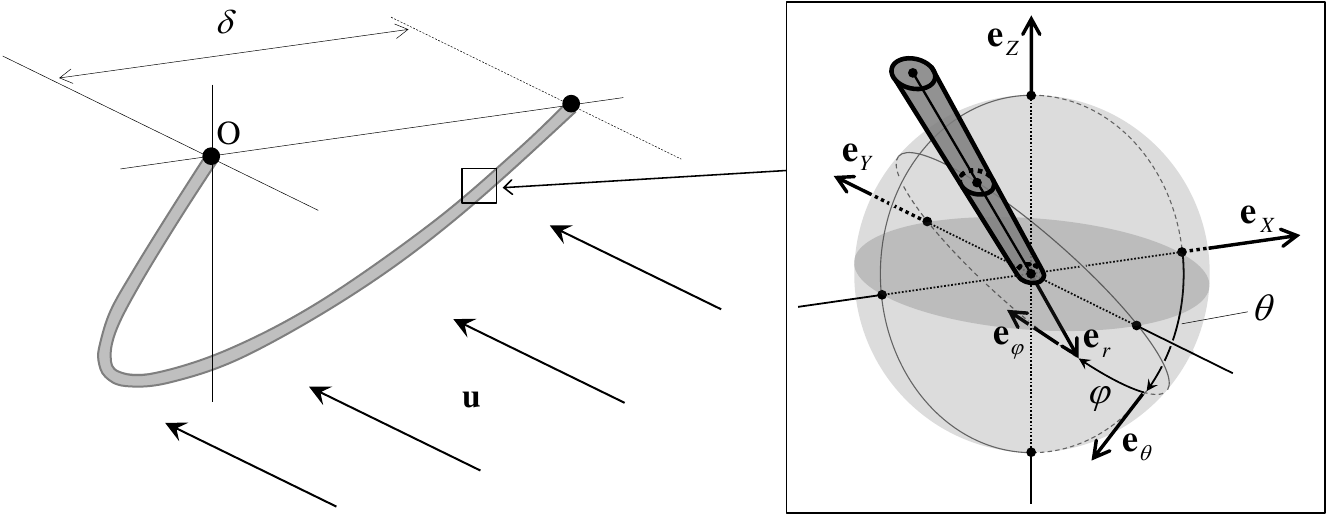}
\caption{VIV of a flexible cable. The inset shows the definition of the local orthonormal basis $(\myvec{e}_r,\myvec{e}_\theta,\myvec{e}_\varphi)$.}
\label{fig:geometry_3d_cable}
\end{center}
\end{figure}

In this work, we consider a long inextensible cable of circular cross-section with density $\rho_S$, length $L$ and diameter $D$ with a large aspect ratio $\Lambda=L/D\gg 1$, see Fig.~\ref{fig:geometry_3d_cable}; its extremities $O$ and $O'$ are fixed and aligned along $\myvec{e}_X$ at a distance $\Delta L$ from each other. By ``cable'' (or equivalently ``string'') we mean that the structure has no bending stiffness, thus the only structural force is the tension force which acts as a Lagrangian multiplier that enforces the inextensibility condition. The reader is referred to Ref.~\mycit{audoly2010elasticity} for a thorough description of the string model. The cable  is immersed in a uniform horizontal cross flow $\myvec{u}=u\myvec{e}_Y$.

The dynamics of the cable are the result of a balance between the cable's inertia, the internal tension, gravity and buoyancy, and flow forces. We focus on vortex-induced vibrations (VIV) for which the cable undergoes small oscillations about a steady mean position as a result of the unsteady vortex shedding on the structure. 

Flow forces on the moving structure can be modelled as the superposition of four different components \mycit{blevins1984handbook,blevins1990}: (i) a drag force $\frac{1}{2}\rho C_D D|\myvec{u}_\perp^r|\myvec{u}^r_\perp$, (ii) a friction force $\frac{\pi}{2}\rho C_F D|\myvec{u}^r|\myvec{u}^r_\parallel$, (iii) an added mass and (iv) an unsteady lift force resulting from periodic vortex shedding in the structure's wake. Here, $\rho$ is the mass density of the fluid and $\myvec{u}^r=\myvec{u}-\dot{\myvec{x}}$ is the local relative velocity of the incoming flow to the cable whose position is noted $\myvec{x}(s,t)$, and ($\myvec{u}^r_\parallel$,$\myvec{u}^r_\perp$) are its components along and normal to the cable, respectively. 

In the following, the problem is written in non-dimensional form by choosing $L$, $1/(2\pi f)$ and $\frac{1}{2}\rho C_D U^2 DL$ as characteristic length, time and force, respectively. %In VIV, the cable's displacement about its mean position is small and is scaled by $D$ in the following.

\subsection{Mean Position of the Structure \label{sec:model_mean_static_position}}
We first focus on the mean (i.e. time-averaged) steady position of the cable, obtained by balancing the effects of the steady fluid forces, the cable's weight, buoyancy and the internal tension (inertia, added-mass and wake effects only contribute to the unsteady dynamics). A local orthonormal basis $(\myvec{e}_r,\myvec{e}_\theta,\myvec{e}_\varphi)$ is used with $\myvec{e}_r=\partial\mean{\myvec{x}}/\partial s$ the local direction of the cable where $\mean{\myvec{x}}$ is the mean position of the cable, and $\myvec{e}_\theta\cdot\myvec{u}=0$ (Fig.~\ref{fig:geometry_3d_cable}). The equilibrium writes

\begin{align}
\left(\chi\,\myvec{e}_r\right)' + \beta\,\sin\varphi\,\,\myvec{e}_r + \cos^2\varphi\,\,\myvec{e}_\varphi - \Gamma\,\myvec{e}_Z= \myvec{0}
\label{eq:static_dless}
\end{align}
\noindent where $0\leq s\leq 1$ is the dimensionless curvilinear coordinate along the cable and $(\cdot)'$ denotes the derivative with respect to $s$. The angles $\theta$ and $\varphi$ determine the local orientation of the local basis with respect to fixed axes (see Fig.~\ref{fig:geometry_3d_cable} and Appendix~\ref{sec:axes}), and $\chi(s)=T(s)/(\frac{1}{2}\rho C_DU^2D)$ is the non-dimensional tension. Equation~\eqref{eq:static_dless}, together with the inextensibility condition $\myvec{e}_r=\partial\mean{\myvec{x}}/\partial s$, is an implicit equation for the mean position $\mean{\myvec{x}}(s)$ and tension $\chi(s)$. Boundary conditions further impose that

\begin{align}
\begin{array}{ccccc}
\int_{0}^{1}\myvec{e}_r\,ds = \delta\,\myvec{e}_X.
\end{array}
\label{eq:static_BC}
\end{align}

The mean problem is thus completely determined by three non-dimensional parameters
\begin{align}
\Gamma=\frac{\pi Dg}{2C_D U^2}\left(\frac{\rho_S}{\rho}-1\right),\quad \beta=\frac{\pi C_F}{C_D},\quad \delta=\frac{\Delta L}{L}.
\end{align}
%SEB: do we really need to introduce this factor \pi for C_F rather than just redefine C_F=\pi*C_F?
%EL : ok with me like it is.

Here $\Gamma$ is a relative measure of the net gravity compared to drag forces, $\beta$ is the ratio of the friction and drag coefficients, and $\delta$ is the relative span. Note that $\delta=1$ corresponds to a straight cable, and decreasing $\delta$ corresponds to a larger sag of the cable in its mean position.

Equations~\eqref{eq:static_dless}--\eqref{eq:static_BC} are solved numerically using an iterative Broyden's method \mycit{broyden1965broydenalgo} for the mean shape and mean tension of the structure. Depending on the value of $\Gamma$, the physical origin of the tension  varies  from flow- to gravity-induced regime. When $\Gamma\ll 1$, gravity and buoyancy effectively balance; as a result of the drag force, the cable is horizontal (``in-flow catenary'') and its tension is due to the sole effect of the fluid (``flow-induced tension''). When $\Gamma\gg 1$, gravity effects are dominant and the cable is vertical (``cross-flow catenary'') while its tension is due to the sole effect of gravity (``gravity-induced tension''). Conversely, when $\Gamma=\bigO{1}$, gravity and fluid effects are comparable in magnitude and they both significantly contribute to the mean position and tension of the cable. 

Whether flow-induced or gravity-induced, the tension within the cable provides it with the ability to carry waves.  To be generic, the typical wave speed should be defined (dimensionally) as $\mathcal{C}=\sqrt{\textrm{max}(T)/m_T}$, with $m_T=\pi D^2(\rho+\rho_S)/4$ the total lineic mass of the cable (including the fluid added mass). This is effectively equivalent to rescaling the non-dimensional tension as $\tilde\chi(s)=\chi(s)/\textrm{max}(\chi)=T(s)/\textrm{max}(T)$, which is done in the rest of the manuscript. We define the characteristic frequency of the waves as $f_0=\frac{\mathcal{C}}{2L}$ which corresponds to the fundamental frequency of a straight cable with uniform tension.

\subsection{Governing Equations for VIV \label{sec:model_VIVs}}

VIV correspond to $\bigO{D}$ displacements of the cable with respect to its mean position in response to the unsteady flow separation in the structure's wake and the resulting unsteady flow forces. These displacements are typically small compared to the structure's length ($D\ll L$) which justifies in the following the linearisation of the equations of motion about the mean position determined above. 

The cable's motion modifies both its position and orientation, with $\bigO{D/L}$ leading order corrections; the position of the cable is thus in non-dimensional form $\mean{\myvec{x}}(s)+ (D/L)\myvec{d}(s,t)$ with $\myvec{d}=\bigO{1}$ the non-dimensional displacement (scaled by $D$). The relative flow velocity and internal tension are also modified both in orientation and in intensity. The non-dimensional tension magnitude in the cable can be decomposed into its steady and unsteady parts as $\tilde\chi(s)+(D/L)\tilde\xi(s,t)$ with $\tilde{\xi}=\bigO{1}$ the non-dimensional fluctuating tension. Linearising the equations of motion about the time-averaged static state determined above leads to the non-dimensional equation for the cable's displacement $\myvec{d}(s,t)$:

\begin{align}
\ddot{\myvec{d}} &= \myvec{f}_{D} + \myvec{f}_{F} + \myvec{f}_{T} + \myvec{f}_{L},
\label{eq:VIV_3d_dless_EOM_cable}
\end{align}
where $\dot{(\,)}$ denotes the time derivative, and the fluctuating drag, friction and tension forces are obtained as
\begin{align}
\myvec{f}_{D}&=  - \frac{\gamma}{\mu}\,\cos\varphi \left( \dot{d}_\theta\,\myvec{e}_\theta + 2\,\dot{d_\varphi}\,\myvec{e}_\varphi \right)\label{eq:VIV_3d_dless_drag}\\
\myvec{f}_{F}
&= - \beta\,\frac{\gamma}{\mu}\,\left(
\dot{d}_r
+ \sin\varphi\,\myvec{e}_Y\cdot\dot{\myvec{d}}
\right)\myvec{e}_r \label{eq:VIV_3d_dless_friction},\\
\myvec{f}_{T}
&= \frac{1}{\pi^2 u^2}\,\left(\tilde\chi'\,\myvec{d}' + \tilde\chi\,\myvec{d}'' + \tilde\xi'\,\myvec{e}_r + \tilde\xi\,\myvec{e}_r'\right)\label{eq:VIV_3d_dless_tension}
\end{align}
with $(d_r,d_\theta,d_\varphi)$ the components of $\myvec{d}$ in the local basis, and $u$, $\gamma$ and $\mu$, respectively defined as the reduced velocity (or frequency ratio), damping and mass ratios~\mycit{facchinetti2004modelVIV,violette2010articleJFS,grouthier2014articleJFS}
\begin{align}
u=\frac{f}{f_0}=2 S_T\frac{L}{D}\frac{U}{\mathcal{C}},\qquad \gamma=\frac{C_D}{4\pi S_T},\qquad \mu=\frac{\pi}{4}\left(1+\frac{\rho_S}{\rho}\right).
\label{eq:gamma_mu_def}
\end{align}

The inextensibility of the cable further imposes that $\myvec{d}'\cdot\myvec{e}_r=0$.

The effect of the unsteady wake on the structure is modelled here as a fluctuating lift force $\myvec{f}_{L}$ which is orthogonal to both the direction of the flow and the axis of the cable. As shown by Franzini et al. \mycit{franzini2009VIVinclinedcylinder}, vortex shedding is mostly governed by the flow orthogonal to the cable's axis and, consequently, the lift force is quadratic in the normal relative flow velocity (i.e. $|\myvec{u}^r_\perp|^2$) with a local fluctuating lift coefficient $C_{L_0}q(s,t)/2$ (with $C_{L_0}$ the lift coefficient of a still cylinder \mycit{facchinetti2004modelVIV}), so that in Eq.~\eqref{eq:VIV_3d_dless_EOM_cable}, $\myvec{f}_L$ is given by
\begin{align}
\myvec{f}_{L}= M\,q(s,t)\,\cos^2\varphi\,\,\myvec{e}_\theta,\qquad \textrm{with    }M=\frac{C_{L_0}}{16\pi^2\mu S_T^2}\cdot
\label{eq:VIV_3d_dless_lift}
\end{align}

So-called wake oscillator models describe the dynamics of this fluctuating lift as a nonlinear van der Pol oscillator forced by the motion of the structure in order to account for the feedback coupling of solid motion on vortex shedding, which is an essential ingredient to lock-in. Previous studies have shown that a local inertial coupling leads to a good agreement with experimental and numerical studies on rigid and flexible structures in VIV~\mycit{facchinetti2004modelVIV,facchinetti2004waveincable,violette2007articleCS,violette2010articleJFS}. The dynamics of the wake variable $q(s,t)$ are then governed by
\begin{align}
\begin{split}
\ddot{q} &+ \epsilon\,\cos\varphi\,\left(q^2-1\right)\,\dot{q}+\cos^2\varphi\,q = A\,\ddot{d_\theta}.
\end{split}
\label{eq:VIV_3d_dless_EOM_wake}
\end{align}
The values of the non-dimensional parameters $\epsilon$ and $A$ of the wake oscillator model need to be determined using experimental (or numerical) data. More details about this model for the wake effects and validation against experimental data for the VIV of rigid cylinders and straight strings are provided in \mycit{facchinetti2004modelVIV,violette2010articleJFS}.\\

In this work, unless stated otherwise, the following numerical values are used for the parameters of the model: $S_T=0.17$, $C_{L0}=0.61$, $C_D=2.0$, $C_F=0.083$, $\epsilon=0.3$ and $A=12$ (e.g. \mycit{facchinetti2004modelVIV}). Two different values of $\mu$ are used in the paper: $\mu=2.79$, which is consistent with previous works on the topic (e.g.  \mycit{facchinetti2004modelVIV,violette2010articleJFS,grouthier2014articleJFS}), and $\mu=\pi/2\approx 1.57$, which corresponds to a neutrally buoyant cable.\\
%SEB: there should be a reference for the choice of S_T, etc...

The extremities of the cable are fixed, which yields the boundary conditions $\myvec{d}|_{s=0}=\myvec{d}|_{s=1}=\myvec{0}$. Equations~\eqref{eq:VIV_3d_dless_EOM_cable}--\eqref{eq:VIV_3d_dless_tension}, \eqref{eq:VIV_3d_dless_lift} and \eqref{eq:VIV_3d_dless_EOM_wake} are numerically integrated in time starting from an initial state where the cable is at rest in its mean position ($\myvec{d}(t=0)=\mathbf{0}$) while the wake is given a small perturbation (at $t=0$, $q=q_0(s)\ll 1$ and $\dot{q}=0$) that triggers the oscillations of the system which eventually reaches a steady oscillatory regime that does not depend on the choice of initial conditions.

%Note that two different values of $\mu$ are used throughout paper: $\mu=2.79$, which is consistent with previous works on the topic (e.g.  \mycit{facchinetti2004modelVIV,violette2010articleJFS,grouthier2014articleJFS}), and $\mu=\pi/2\approx 1.57$, which corresponds to a neutrally buoyant structure.

\subsection{Modeling Energy Harvesting}

\begin{figure}
\begin{center}
\includegraphics[scale=1.00]{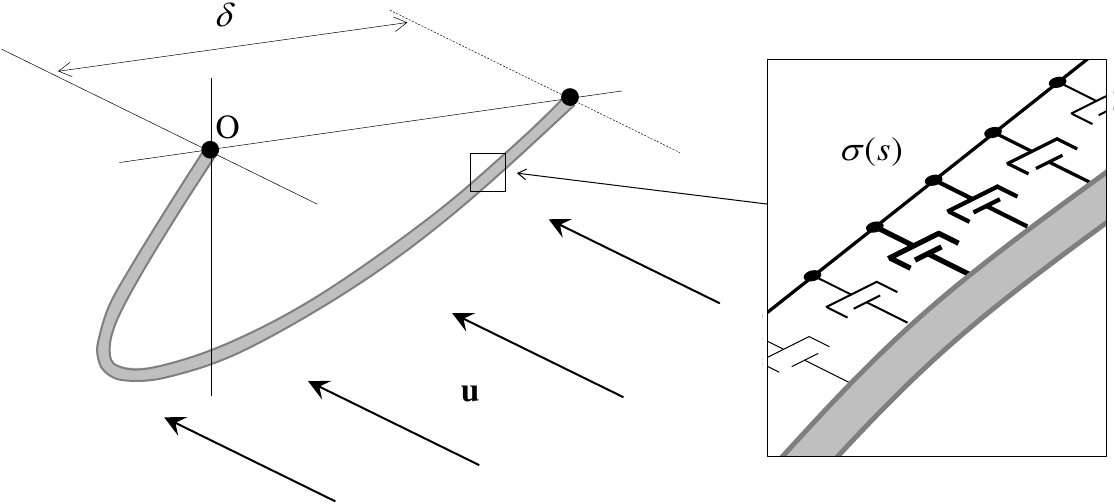}
\caption{Energy harvesting of a flexible cable in VIV using a continuous distribution of harvesting devices (i.e. dampers) of local intensity $\sigma(s)$. For the sake of simplicity, we draw a single line of dampers in the figure.}
\label{fig:geometry_3d_cable_w_distribdamp}
\end{center}
\end{figure}

This work focuses on the possibility of harvesting energy from a flow using the cable's VIV, and on the optimal strategy to maximize the energy output. Energy harvesting amounts to converting some of the energy associated with the vibrations into a usable form -- typically electricity. As it effectively removes some mechanical energy from the structure, energy harvesting must be modelled explicitly in order to properly account for its effect on the vibration itself. The simplest model is that of a pure linear damping force. This damping can be either distributed (i.e. present all along the structure) or localized at some position along the cable. The former will be our initial focus, and corresponds to an additional non uniform damping force $\myvec{f}_{H}$ in the equations of motion of the cable, Eq.~\eqref{eq:VIV_3d_dless_EOM_cable}:

\begin{align}
\begin{split}
\myvec{f}_{H}
&= -\sigma(s)\,\dot{\myvec{d}}.
\end{split}
\label{eq:VIV_3d_dless_distribharvester}
\end{align}
The damping intensity $\sigma(s)$ (scaled by $2\pi f m_T$) may depend on the position $s$ (see Fig.~\ref{fig:geometry_3d_cable_w_distribdamp}), with the requirement that $\sigma(s)\geq 0$ (passive energy harvester). The latter (i.e. localized harvesting) will be explored subsequently.
In both cases, the power $\mathcal{P}_{\textrm{el}}$ dissipated in the dampers acts as a proxy  for the power harvested from the fluid flow by the device. Its efficiency is therefore defined as the ratio of $\mathcal{P}_{\textrm{el}}$ to the reference kinetic energy flux through the reference surface $\mathcal{A}=LD$

\begin{align}
\eta=\frac{\mathcal{P}_{\textrm{el}}}{\frac{1}{2}\rho U^3 \mathcal{A}}=16\,\mu\,\pi^3\,S_T^3\,\,\int_{0}^{1}\sigma(s)\left\langle\left\|\dot{\myvec{d}}\right\|^2\right\rangle ds,
\label{eq:VIV_3d_distrib_efficiency}
\end{align}
with $\langle\cdot\rangle$ the time average, computed once the system reaches a permanent saturated regime. Note that $\eta$ can also be understood as the non-dimensional power output.

\section{Optimal Energy Harvesting Strategy  \label{sec:opti_harvest_strategy_3d}}

Finding the optimal harvesting strategy, i.e. determining where the vibration energy of the cable must be harvested in order to maximize the output power, is equivalent to determining the function $\sigma(s)$ that maximizes the harvesting efficiency $\eta$ defined in Eq.~\eqref{eq:VIV_3d_distrib_efficiency}.  The location of maximal vibration depends on the vibration mode selected by the flow velocity (see the experiments of Ref.~\mycit{king1995modeslockin}). The optimal harvesting strategy $\sigma_\text{opt}(s)$ and the corresponding efficiency $\eta_\text{opt}$ are therefore also expected to depend on the reduced flow velocity $u$.

\subsection{Optimal Harvesting from VIV of Straight Cables}

We first focus on the simplest geometry, namely that of a straight cable with uniform tension ($\delta=1$). In this special case, the only non-zero component of the displacement is $d_\theta=-z$ and the coordinate $s$ matches $x$.

In that limit, the equations governing the VIV dynamics take a much simpler form, which is amenable to a full gradient-based optimization using adjoint methods to compute the gradient of the harvesting efficiency with respect to $\sigma(s)$. 

Adjoint methods represent a popular and powerful approach to optimization problems involving partial differential equations, e.g. see \mycit{cao2003adjoint}. Here, we follow an approach similar to \mycit{meliga2014adjointgradient}: the gradient $\nabla_\sigma [\eta](s)$ of the efficiency with respect to the damping distribution $\sigma(s)$ is a function of $s$ and is computed as
\begin{align}
\nabla_\sigma [\eta](s)=16\,\mu\,\pi^3\,S_T^3\,\left\langle \dot{z}\left(\dot{z}-z^\dag\right)\right\rangle
\label{eq:gradient}
\end{align}
where $z^\dag$ is the adjoint displacement. The derivation of the adjoint equations for $(z^\dag, q^\dag)$ from the direct equations, Eqs.~\eqref{eq:VIV_3d_dless_EOM_cable} and \eqref{eq:VIV_3d_dless_EOM_wake}, is presented in Appendix~\ref{sec:adjoint}. Starting from an initial guess for $z(s,t)$ and $q(s,t)$ without any damping ($\sigma=0$), the direct and adjoint equations are solved at each iteration to obtain $\nabla_\sigma[\eta]$ using Eq.~\eqref{eq:gradient}; a steepest-ascent algorithm is then used until convergence to the optimal damping distribution $\sigma_{\textrm{opt}}(s)$.

\begin{figure}
\begin{center}
\includegraphics[width=.85\textwidth]{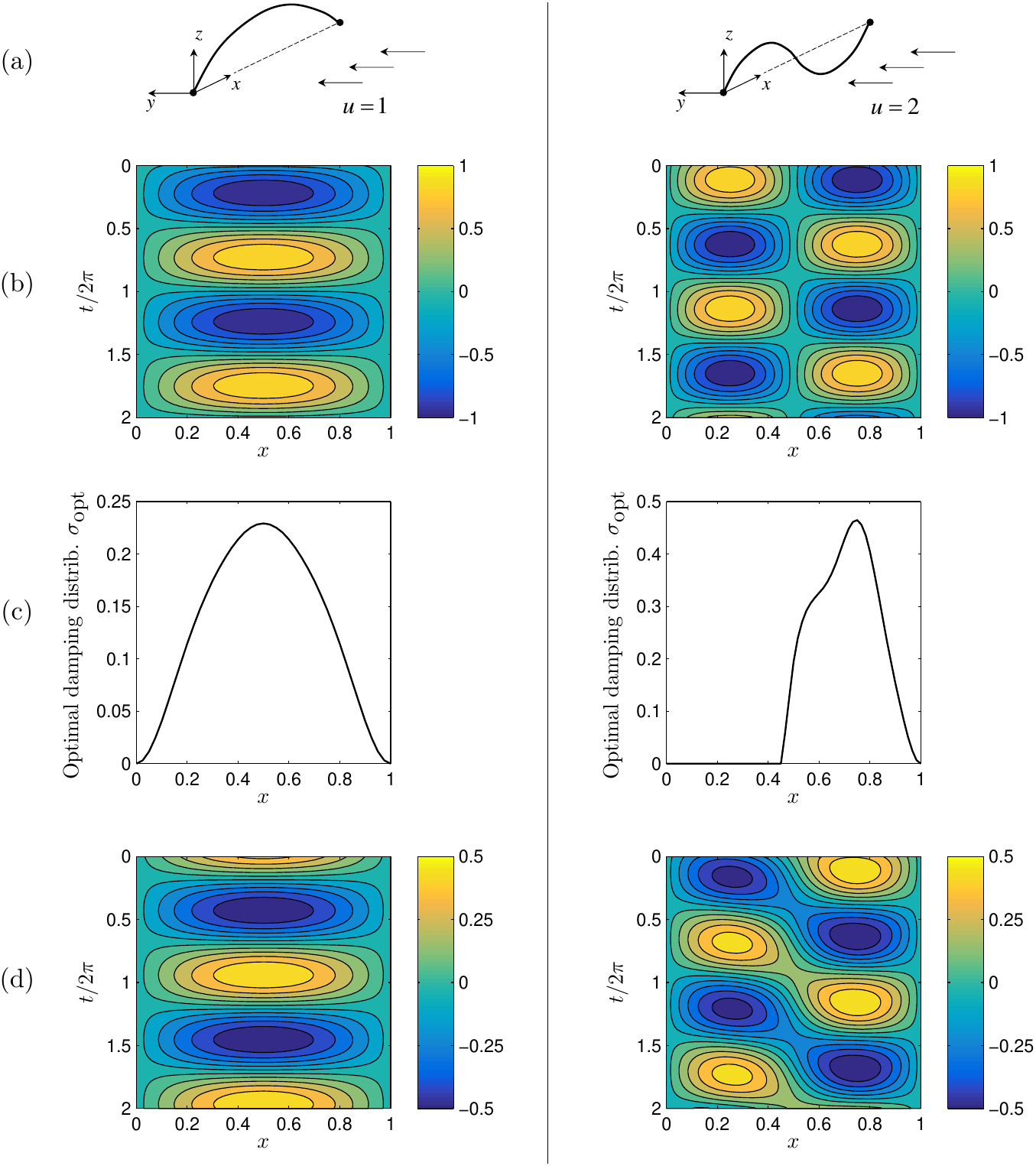}
\caption{Optimization of the harvesting efficiency for VIV in modes one (left) and two (right) of a straight cable ($\delta=1$) with $\mu=2.79$. From top to bottom: (a) sketch of the cable, (b) contours of the cable deflection $z$ without energy harvesting ($\sigma=0$), (c) optimal damping distribution $\sigma_{\textnormal{opt}}$, and (d) contours of $z$ with $\sigma=\sigma_{\textnormal{opt}}$. }
\label{fig:VIV_straight_distrib_opti_modes12}
\end{center}
\end{figure}

In VIV of flexible structures, the flow velocity selects and excites the eigenmode of the structure whose frequency is the closest to the Strouhal frequency (see experiments in \mycit{king1995modeslockin}). For the wake oscillator model used here, linear stability analysis indicates that the $n$-th mode of the straight cable is specifically excited when $u\approx n$~\mycit{violette2010articleJFS}. For instance, when $u=1$, the cable deforms in mode one and $\sigma_\text{opt}(s)$ is expected to be large at the center of the cable, where the kinetic energy is the largest. When $u=2$ however, the cable deforms in mode two and there is little to no available energy at the center of the cable; thus the optimal harvesting strategy in mode one is unlikely to carry over to mode two, and the optimal harvesting strategy must be determined for each individual mode by setting $u$ to the appropriate value. Figure~\ref{fig:VIV_straight_distrib_opti_modes12} presents the optimization results for a straight cable in mode 1 and 2.

For $u=1$ (mode-1 lock-in), the distribution of deflection $z$ is qualitatively similar with no damping (initial guess) and with an optimal damping (final result). The amplitude of the oscillations with the optimal harvesting strategy is nearly half that of the oscillations without energy harvesting, indicating a significant energy extraction from the oscillations. The optimal damping distribution $\sigma_\text{opt}$ respects the mode one symmetry.

For $u=2$ (mode-2 lock-in), the distribution of deflection $z$ without damping (initial guess) corresponds to a mode-two deformation -- which validates our criterion $n=u$. This is not the case for the $z$-contours in the optimal-damping situation: the optimal strategy corresponds to a non-symmetric damping distribution $\sigma_\text{opt}(s)$ (Fig.~\ref{fig:VIV_straight_distrib_opti_modes12}), for which most of the energy extraction occurs on one half of the cable. Progressive waves carry energy from the undamped part of the cable to the harvesting location. The optimal efficiency is similar for $u=1$ and $u=2$, with $\eta_{\textrm{opt}}=8.1\%$ and $\eta_{\textrm{opt}}=8.6\%$, respectively.

These results are in fact generic: for any $n\geq 2$, the optimization algorithm  leads to an optimal strategy with harvesting concentrated on a reduced fraction of the cable near one of its fixed ends, while the rest of the cable is undamped, and energy is carried to the harvesting region by travelling waves along the cable.

Those results are particularly relevant from a practical and engineering point of view: regardless of the selected mode except mode 1 (and therefore for any sufficiently large velocity), the optimal strategy consists in restricting the harvesting system to a limited fraction of the cable near one of its attachment points, rather than distributing it along the entire cable.

\subsection{Optimal Harvesting from VIV of Catenary Cables}

We can extend the previous approach and results to the general configuration of Fig.~\ref{fig:geometry_3d_cable_w_distribdamp}. The optimal harvesting strategy $\sigma_\text{opt}(s)$ now depends on three parameters: $\delta$ and $\Gamma$, that set the mean position of the cable, and $u$, which sets the vibration mode of the cable excited by the flow. Assuming a piecewise constant damping $\sigma(s)$, the damping distribution $\sigma_\text{opt}(s)$ is computed for different values of $(\delta,\Gamma,u)$ using a steepest-ascent algorithm, now computing the gradient numerically. % SEB: is that how you do it ? (direct numerical calculation of gradient)

\begin{figure}[h!]
\begin{center}
\includegraphics[width=.43\textwidth]{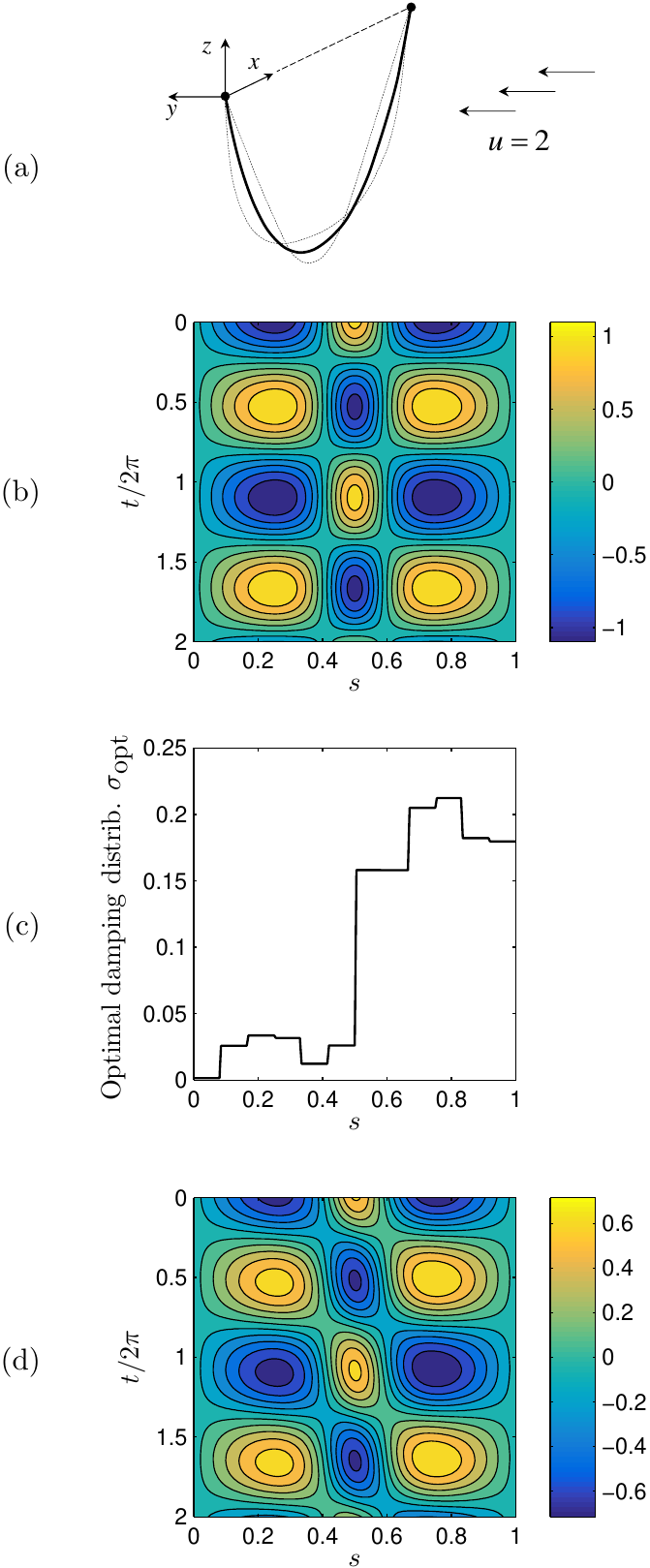}
\caption{Optimization of the harvesting efficiency for a mode 2 VIV ($u=2$) of a cross-flow catenary ($\Gamma\gg 1$) with  $\delta=0.5$ and $\mu=2.79$. (a) sketch of the cable in mode 2 lock-in, (b) contours of $d_\theta$ without energy harvesting ($\sigma=0$), (c) optimal damping distribution $\sigma_{\textnormal{opt}}$, (d) contours of $d_\theta$ with $\sigma=\sigma_{\textnormal{opt}}$. }
\label{fig:VIV_3d_heavy50_distrib_opti_mode2}
\end{center}
\end{figure}

Figure~\ref{fig:VIV_3d_heavy50_distrib_opti_mode2} shows the optimal distribution of damping and resulting dynamics obtained for a cross-flow catenary cable ($\eta_{\textrm{opt}}=7.6\%$). As for the straight cable, the optimal damping distribution is not symmetric and waves are observed to carry energy toward the harvesting site. This behaviour is in fact observed for all geometries provided the excited mode number is greater than 2. This generality therefore suggests to go one step further and investigate the performance of harvesting energy at a single point located \emph{at} the extremity of the cable rather than in its vicinity. 

\subsection{Local Point-Wise Damper as Optimal Harvesting Strategy \label{sec:opti_strat_is_local_damper}}
\begin{figure}
\begin{center}
\includegraphics[scale=1.00]{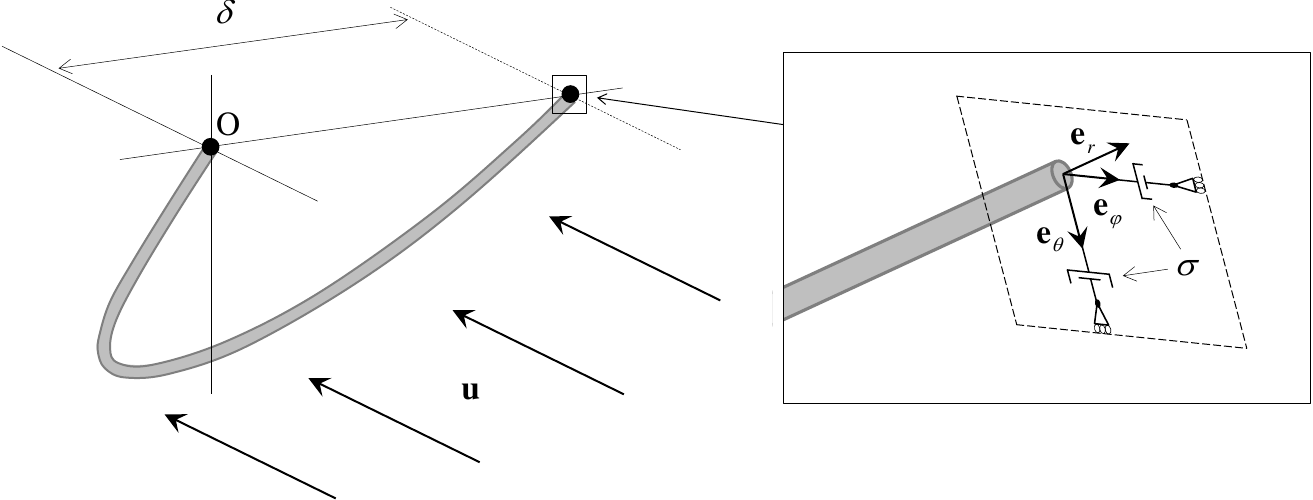}
\caption{Energy harvesting of a flexible cable in VIV using a local harvester of intensity $\sigma$ at $s=1$.}
\label{fig:geometry_3d_cable_w_local_harvester}
\end{center}
\end{figure}
To this end, the fixed boundary condition at $s=1$ is now relaxed to allow for transverse displacements of the cable and energy extraction. More precisely, at $s=1$, the cable can slide in the local ($\theta$,$\varphi$) plane and energy harvesting is modelled as a linear viscous force resisting its velocity with a damping intensity $\sigma$ (Fig.~\ref{fig:geometry_3d_cable_w_local_harvester}). Note that $\sigma$ is a scalar value here. The cable and wake dynamics are still governed by Eqs.~\eqref{eq:VIV_3d_dless_EOM_cable} and \eqref{eq:VIV_3d_dless_EOM_wake} (without an additional damping force) and the boundary condition at $s=1$ now balances local tension and viscous forces (the fixed boundary condition at $s=0$ remains unchanged):

\begin{align}
\begin{array}{ccc}
d_r=0\,, & \myvec{d}'\cdot\myvec{e}_\theta+\pi^2u^2\,\sigma\,\dot{d_\theta}=0\,, & \myvec{d}'\cdot\myvec{e}_\varphi+\pi^2u^2\,\sigma\,\dot{d_\varphi}=0
\end{array}
\label{eq:VIV_3d_local_BC1_dless}
\end{align}
The harvesting efficiency of this system is still defined as the ratio of the average power dissipated in the damper to the kinetic energy flux through the area occupied by the cable:
\begin{align}
\eta = 16\,\mu\,\pi^3\,S_T^3\,\sigma\,\left\langle\left\|\left.\dot{\myvec{d}}\right|_{s=1}\right\|^2\right\rangle
\label{eq:VIV_3d_local_efficiency}
\end{align}
\noindent The optimization now consists in maximizing $\eta$ with respect to the scalar parameter $\sigma$. 

\begin{figure}
\begin{center}
\includegraphics[scale=1.00]{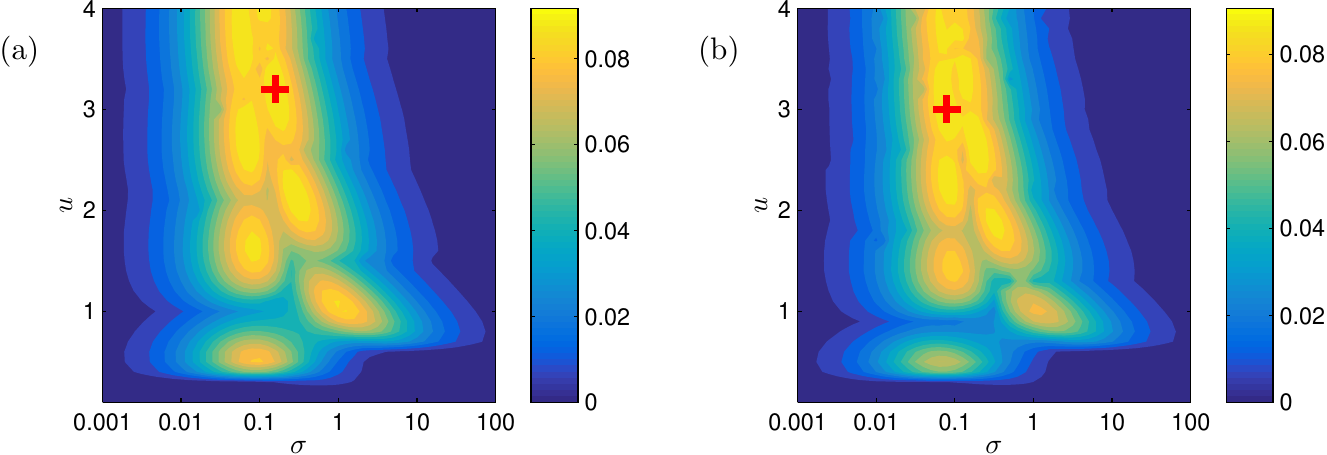}
\caption{Efficiency $\eta$ as a function of the damping $\sigma$ and the reduced velocity $u$ for (a) the straight cable and (b) the cross-flow catenary with span $\delta=0.5$. Red crosses are placed at the maximum efficiency. (a) $\eta_\text{opt}=9.2\%$, $u_\text{opt}=3.2$, $\sigma_\text{opt}=0.16$, and (b) $\eta_\text{opt}=9.1\%$, $u_\text{opt}=3.0$, $\sigma_\text{opt}=0.080$. For both contours, $\mu=2.79$.}
\label{fig:VIV_efficiency_contour_straight_heavy50}
\end{center}
\end{figure}

For a given geometry (i.e. given $\delta$ and $\Gamma$), the optimal damping strategy and efficiency are determined by directly computing $\eta(u,\sigma)$ and finding its absolute maximum. Figure~\ref{fig:VIV_efficiency_contour_straight_heavy50} shows the evolution of $\eta(u,\sigma)$ for the configurations of Figs.~\ref{fig:VIV_straight_distrib_opti_modes12} (straight cable) and \ref{fig:VIV_3d_heavy50_distrib_opti_mode2} (catenary).

The maximum efficiency for the straight cable and the cross-flow catenary is slightly more than 9\%, which exceeds the maximum performance obtained with distributed harvesters: the local harvesting strategy is therefore optimal for both configurations. This finding is again generic, and valid for most geometries: for $\delta$ ranging between 0.25 and 1 and $\Gamma \ll 1$, $\Gamma \sim 1$ and $\Gamma \gg 1$ we find that the optimal efficiency of a local damper is equivalent or slightly higher than the optimal efficiency of distributed damping. 

We note that the contours of Fig.\ref{fig:VIV_efficiency_contour_straight_heavy50} are qualitatively similar to the efficiency map obtained by Grouthier et al.~\mycit{grouthier2014articleJFS} with a hanging cable attached to a local harvester at its upper end. The agreement is also quantitative if we account for the different values for the lift coefficient $C_{L0}$ used in the works (0.63 in the present work, 0.80 in \mycit{grouthier2014articleJFS}). Those similarities suggest that the influence of the detailed geometry on the performance of cables with gravity-induced tension is weak. This is discussed further in Section~\ref{sec:discuss}\ref{sec:disc_robust_crossflow}. Besides, we note that the periodic cable studied in \mycit{grouthier2014articleJFS} exhibits the same maximum efficiency as the hanging cable and catenary cables although its efficiency map is considerably different.

\section{Peak performance and robustness \label{sec:discuss}}
Energy harvesting using VIV was originally proposed using a rigid cylinder system (VIVACE system \mycit{bernitsas2008vivace}). The goal of this section is to compare the overall performance of the flexible cable system in  Fig.~\ref{fig:geometry_3d_cable_w_local_harvester} to this reference configuration, with a particular focus on its robustness to variability in the flow velocity.

\subsection{Harvesting performance and limitation of rigid cylinders in VIV}

\begin{figure}
\begin{center}
\includegraphics[scale=1.00]{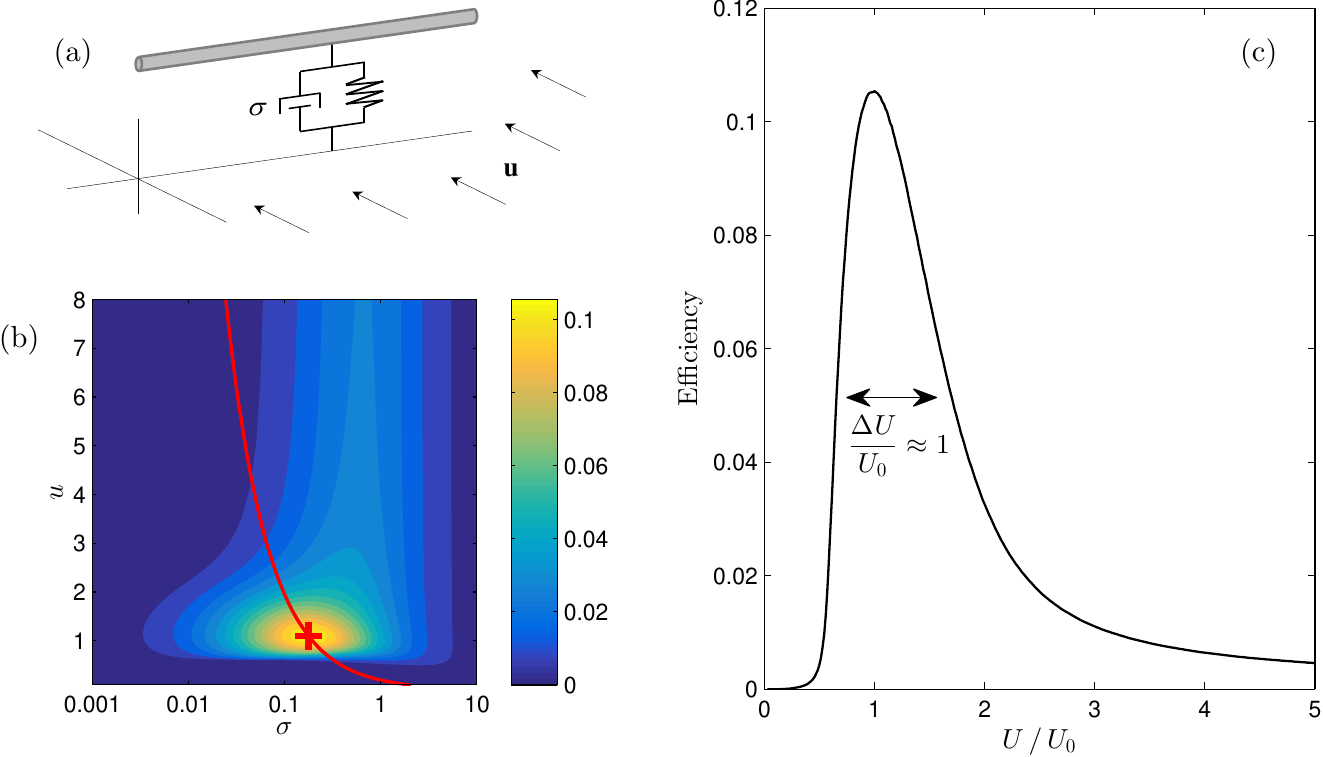}
\caption{Energy harvesting by a rigid cylinder with $\mu=2.79$. (a) is sketch of the system, (b) is the harvesting efficiency $\eta$. The red cross is the optimal configuration ($\eta_\text{opt}=10.5\%$ with $u_\text{opt}=1.1$ and $\sigma_\text{opt}=0.18$), and the red line corresponds to variations of the dimensional flow velocity for an optimal system (all other quantities being fixed). (c) shows the evolution of the efficiency with flow velocity around the optimal configuration (red line in (b)).
}
\label{fig:VIV_efficiency_contour_rigid}
\end{center}
\end{figure}

Ref.~\mycit{grouthier2014articleJFS} carried out a detailed study of the rigid cylinder system, and we present only a brief summary here, together with an update of the numerical results (Fig.~\ref{fig:VIV_efficiency_contour_rigid}) for the particular parameter values considered here (in particular, the Strouhal number and lift coefficient are modified in comparison with Ref.~\mycit{grouthier2014articleJFS}). This system extracts energy from VIV of a rigid cylinder and it is the origin of the VIVACE concept.

The system is a rigid cylinder mounted horizontally on an elastic foundation (stiffness $k$ per unit length) and attached to an energy harvester of intensity $R$ (per unit length). Its resonance frequency in still flow is therefore $f_0=\sqrt{k/m_T}/2\pi$. It is immersed in a horizontal cross-flow $U$ and its motion is restricted to vertical vibrations along $\myvec{e}_Z$. The harvesting efficiency $\eta$ (defined as above) depends solely on the reduced velocity (or frequency ratio) $u=f/f_0$ and damping intensity $\sigma=R/(2\pi f m_T)$.

Optimizing the design of such a system, which is meant to be deployed in a current of typical mean velocity $U_0$, amounts to choosing its dimensional characteristics such that $u=u_\text{opt}$ and $\sigma=\sigma_\text{opt}$ under normal conditions ($U=U_0$) so that $\eta=\eta_{\textrm{opt}}=10.5\%$. The velocity of actual geophysical flows typically exhibits a significant variability in magnitude (in particular for tidal currents). All other quantities being fixed (e.g. geometry, stiffness,...), a change in the dimensional flow velocity $U$ modifies $(u,\sigma)$ and therefore induces a departure from the optimal operating conditions and a reduction of the system's efficiency~(Fig.~\ref{fig:VIV_efficiency_contour_rigid}, c). This reduced efficiency results from a departure from the optimal lock-in conditions: when $U\neq U_0$, the shedding frequency $f$ differs from the fundamental frequency of the cylinder.

 The efficiency peak around $U_0$ is narrow: its width at half height, i.e. the range of $U$ in which the efficiency is at least half its peak value, is almost equal to $U_0$. Thus, the system is efficient only when $0.67 \leq (U/ U_0) \leq 1.7 $. In a practical situation, this could result in a negligible efficiency of the system for a significant fraction of its operating time.

Below, we show that extracting energy from VIV of flexible cables rather than rigid cylinders provides a simple and efficient method to remedy this issue, while still maintaining comparable peak performance.

\subsection{Harvesting performance and robustness of flexible cables in VIV}

We now turn back to the case of a flexible cable with a local harvester at $s=1$ (Fig.~\ref{fig:geometry_3d_cable_w_local_harvester}). The performance of this system as well as its sensitivity to variations in the flow velocity depend on the shape of the cable determined by the span $\delta$ and reduced gravity $\Gamma$. The performance of the flexible cable with respect to $u$ and $\sigma$ is analysed for $0.25\leq \delta\leq 1$ and $\Gamma\ll 1$ (in-flow catenary) and $\Gamma\gg 1$ (cross-flow catenary) by computing an efficiency map similar to that obtained for the rigid cylinder in Fig.~\ref{fig:VIV_efficiency_contour_rigid}-b or those obtained for cables in Fig.~\ref{fig:VIV_efficiency_contour_straight_heavy50}. The optimal harvesting efficiency $\eta_\text{opt}$ of each particular geometry is then determined (together with the corresponding optimal parameters $u_{\textrm{opt}}$ and $\sigma_{\textrm{opt}}$), as well as the influence on $\eta$ of a change in the dimensional velocity $U$ away from its design value $U_0$. 

\begin{figure}
\begin{center}
\includegraphics[scale=1.00]{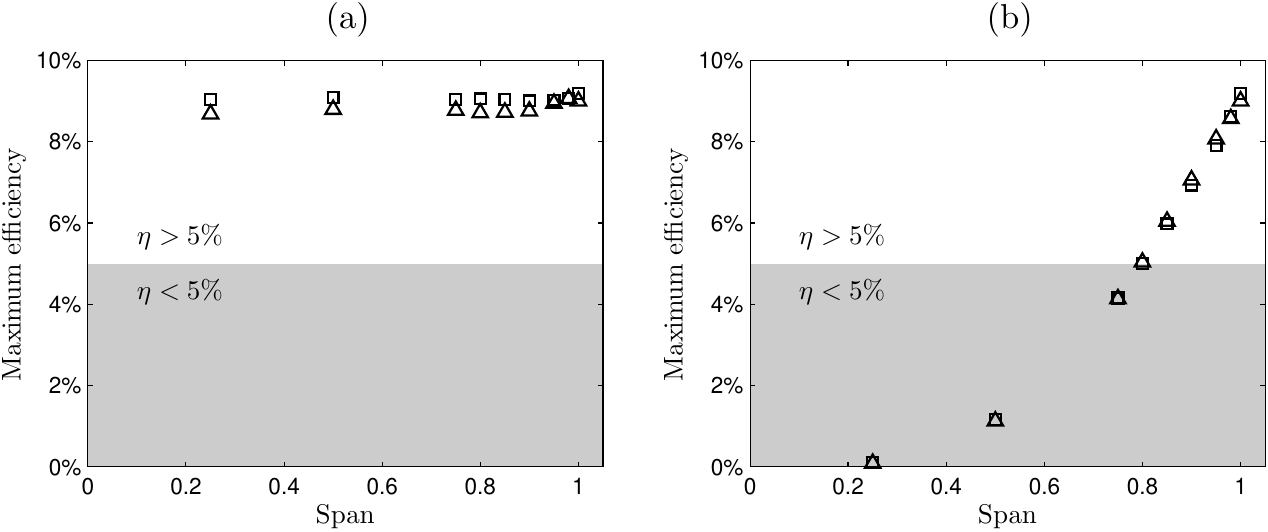}
\caption{Maximum harvesting efficiency as a function of the span $\delta$ for (a) cross-flow catenaries and (b) in-flow catenaries. Key to the plots: {\tiny $\square$} for $\mu=2.79$, {\tiny $\triangle$} for $\mu=\pi/2$.}
\label{fig:VIV_max_efficiency_vs_span_for_heavy_and_nb}
\end{center}
\end{figure}

The performance and robustness of the system to flow fluctuations is characterized by two important quantities: the peak efficiency $\eta_{\textrm{opt}}$ and the peak width $\Delta U/U_0$ measuring the relative velocity range over which the efficiency is at least $\eta_{\textrm{opt}}/2$.
Both quantities are critical: the former characterizes the efficiency of the system and the latter its robustness to flow velocity variability. They are respectively reported in Fig.~\ref{fig:VIV_max_efficiency_vs_span_for_heavy_and_nb} and in Table~\ref{tab:VIV_efficiency_peak_width}.

\begin{table}
\vspace{-7pt}
\centering
{\renewcommand{\arraystretch}{1.2}
\begin{tabular}{|c|c|c|c|c|c|c|}
\cline{2-7}
\multicolumn{1}{c|}{} & \multicolumn{3}{c|}{$\mu=\pi/2$} & \multicolumn{3}{c|}{$\mu=2.79$}
\\
\cline{2-7}
\multicolumn{1}{c|}{} & $\delta=1$ & $\delta=0.75$ & $\delta=0.5$ & $\delta=1$ & $\delta=0.75$ & $\delta=0.5$
\\
\hline
\minitab[c]{In-flow catenary $\Gamma\ll 1$\\\includegraphics[scale=1.00]{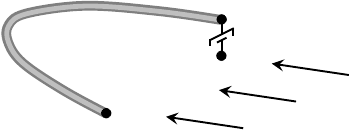}} & 6.9 & 7.8 & 2.5 & 8.2 & 3.1 & 2.3
\\
\hline
\minitab[c]{Cross-flow catenary $\Gamma\gg 1$\\\includegraphics[scale=1.00]{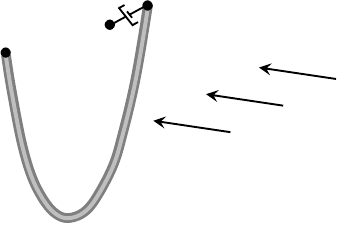}} & 3.3 & 2.6 & 3.1 & 3.6 & 3.2 & 3.6
\\
\hline
\minitab[c]{Rigid cylinder\\\includegraphics[scale=1.00]{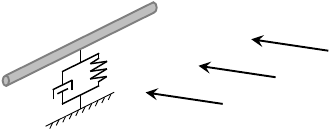}} & \multicolumn{3}{c|}{1.6}  & \multicolumn{3}{c|}{1.0}
\\
\hline
\end{tabular}
\vspace{-4pt}
}
\caption{Width of efficiency peaks at half height for systems extracting energy from VIV of cables and rigid cylinders.}
\label{tab:VIV_efficiency_peak_width}
\end{table}

Efficient energy harvesting can be achieved using cross-flow catenaries ($\Gamma\gg 1$, gravity-induced tension) of any span or in-flow catenaries ($\Gamma\ll 1$, flow-induced tension) that are almost straight ($\delta\approx 1$) as shown in Fig.~\ref{fig:VIV_max_efficiency_vs_span_for_heavy_and_nb}. This efficiency is comparable to the rigid cylinder performance, albeit slightly lower. In-flow catenaries with larger sag only show poor peak performance. 

\begin{figure}[h!]
\vspace{-7pt}
\centering
\includegraphics[scale=1.00]{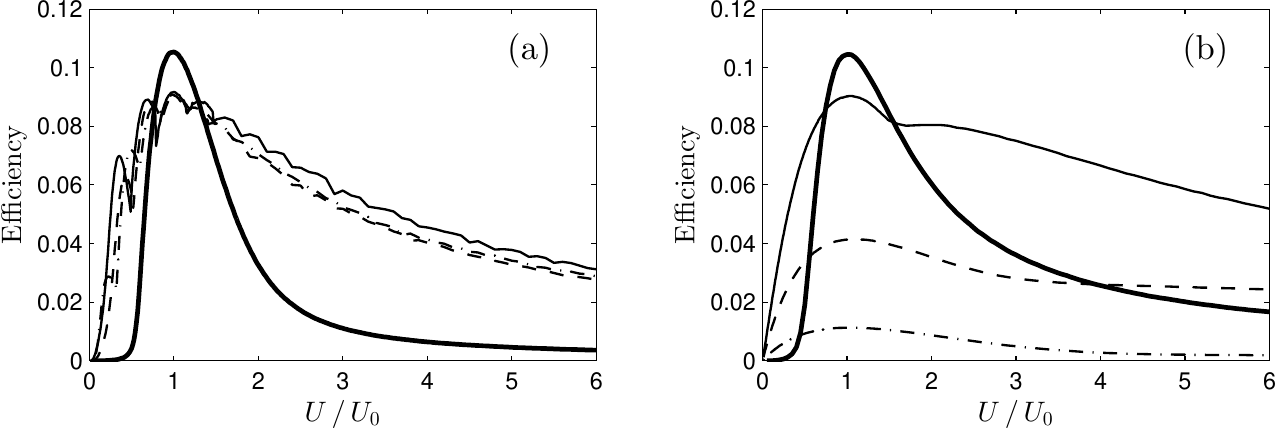}
\caption{Evolution of the harvesting efficiency $\eta$ as a function of the flow velocity $U/U_0$ for (a) rigid cylinders and  cross-flow catenaries with $\mu=2.79$ and (b) rigid cylinders and  in-flow catenaries with $\mu=\pi/2$. $U_0$ denotes the flow velocity corresponding to the optimal design. Key to the plots: rigid cylinder (thick solid), catenary with span 1 (thin solid), catenary with span 0.75 (dashed), catenary with span 0.5 (dash-dotted).}
\label{fig:VIV_efficiency_along_isodamping__rigid_heavy_nb}
\vspace{-5pt}
\end{figure}

All configurations presented are nevertheless significantly more robust to variations of the flow velocity than their rigid counterpart (Table~\ref{tab:VIV_efficiency_peak_width}). In particular, the efficiency peak width of the straight in-flow catenary ($\delta\approx 1$, $\Gamma\ll 1$) is seven times as large as that of the rigid cylinder, which makes it an attractive candidate for an energy harvesting system based on VIV. The comparisons between these different systems is summarized in Fig.~\ref{fig:VIV_efficiency_along_isodamping__rigid_heavy_nb}, by plotting the variation of the efficiency with the flow velocity $U$ around the optimal configuration. 

\subsection{Robustness of Energy Harvesting from VIV of Cross-Flow Catenaries \label{sec:disc_robust_crossflow}}

The mean tension of the cross-flow catenaries is determined solely by gravity, and is thus independent of the flow velocity. Hence, the eigenmodes of the cable have constant frequencies. When the velocity of the flow increases, so does the Strouhal frequency and as as a result higher modes are successively locked-in. The transitions between the different modes of the cable correspond to the kinks in the curves in Fig.~\ref{fig:VIV_efficiency_along_isodamping__rigid_heavy_nb}-a.
 The robustness of cross-flow catenaries to flow velocity variations therefore arises from their ability to adapt their deformation mode to the surrounding flow and remain at lock-in, which the rigid cylinder is unable to do since it only has a single resonance frequency. Noticeably, the value of the relative span $\delta$ of cross-flow catenaries has only a minor impact on their efficiency in terms of optimal performance and robustness to flow velocity variations.

Interestingly, results obtained by \mycit{grouthier2014articleJFS} show that the robustness to flow variations of cross-flow catenaries is comparable to that of hanging cables (width of efficiency peak is 3.6 for the former and 3.7 for the latter). That observation agrees with the minor effect of the span $\delta$ on the performance of cross-flow catenaries in suggesting that the robustness of energy harvesting from a cable with gravity-induced tension is only determined by the mechanism of transitioning between mode lock-in and is independent of the cable shape.

\subsection{Robustness of Energy Harvesting from VIV of In-Flow Catenaries}

For in-flow catenaries, buoyancy and gravity balance and the tension is induced by the flow drag and friction forces. It is therefore set by the flow velocity magnitude and larger flow velocity results in a larger tension in the catenary, that scales as $U^2$. An increase in the flow velocity $U$ now has two consequences: (i) a linear increase in the Strouhal frequency and (ii) a quadratic increase of the tension in the cable, resulting in a linear increase with $U$ of its eigenfrequencies.  Both fluid and solid frequencies increase in the same proportion, and the system passively adapts to remain at lock-in despite the flow variations. The efficiency is therefore only weakly dependent on the dimensional value of the flow velocity $U$ (Fig.~\ref{fig:VIV_efficiency_along_isodamping__rigid_heavy_nb}-b). The robustness of in-flow catenaries to flow velocity variations therefore arises from the passive adaptation of their internal tension to the surrounding flow. Such passive control is not possible with a rigid cylinder.

\section{Conclusion}\label{sec:conclusion}
This work presented a detailed and systematic analysis of the energy harvesting performance of flexible cables in VIV. A fundamental symmetry-breaking in the optimal harvesting strategy was identified that leads to  recommending a concentration of the energy harvesting devices near one of the fixed ends of the cable. We further showed that, out of all distributions possible along the entire cable, the very simplest one performs best, namely a single energy harvester located at one of the extremity of the cable; beyond its fundamental importance, this result also has significant practical engineering implications. 

With similar peak performance, flexible cables in VIV are also significantly more robust than their rigid-cylinder counterpart with respect to variations  in the flow velocity (their range of efficient operation is three to eight times larger). In particular, a fundamental physical insight on this increased robustness was obtained and two different mechanisms were identified: (i) the passive control by the flow velocity of the deformation mode (for systems with gravity-induced tension) and (ii) the passive control of the internal tension of the cable (for systems with flow-induced tension); both maintain the system at lock-in with high energy harvesting efficiency. 

\begin{figure}[h!]
\begin{center}
\includegraphics[width=.8\textwidth]{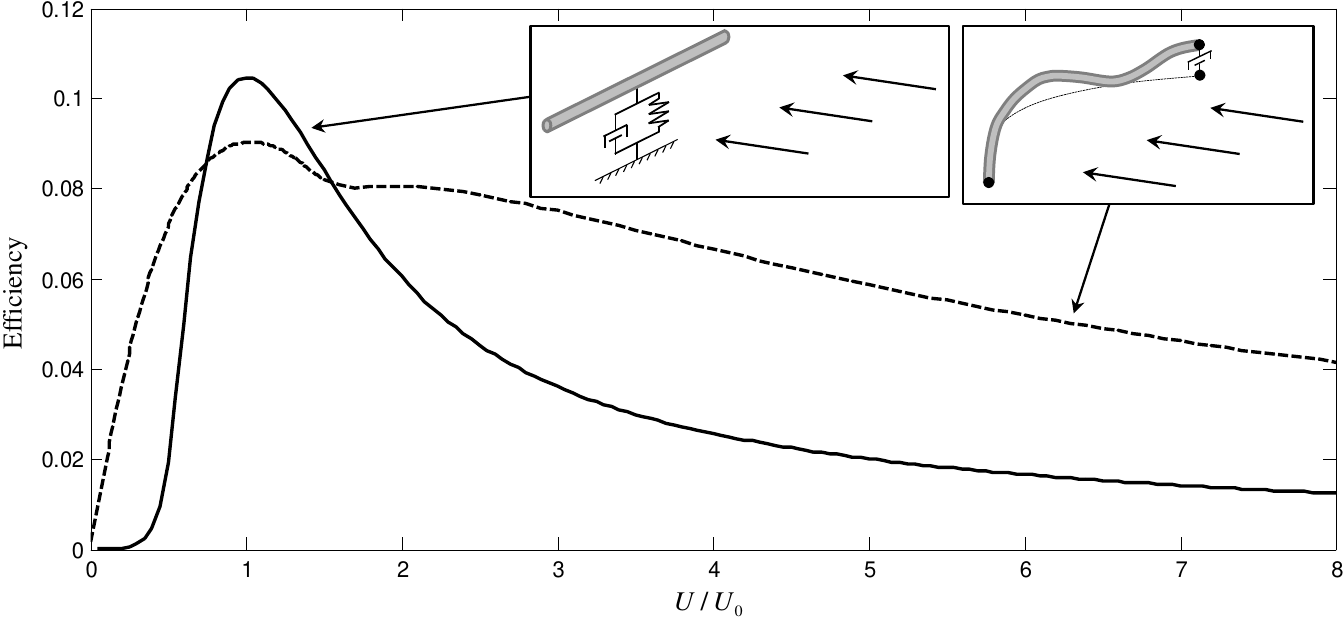}
\caption{Evolution of the harvesting efficiency with the flow velocity for harvesting systems based on VIV of rigid cylinders (solid) and straight cables with flow-induced tension (dashed). $\mu=\pi/2$ for the rigid cylinder and the cable.}
\label{fig:VIV_efficiency_vs_flow_fluctuations_final}
\end{center}
\end{figure}

The best performing design is identified as an almost straight cable (i.e. with minimal sag) with flow-induced tension, which can be practically achieved with a neutrally-buoyant cable. From an engineering point of view, this system is not more difficult to build than a device based on a rigid cylinder, and Fig.~\ref{fig:VIV_efficiency_vs_flow_fluctuations_final} summarizes the comparison of its performance to the rigid-cylinder system. Further, for a neutrally-buoyant cable the direction of gravity is irrelevant, and the system can be fixed vertically, providing an interesting design solution to passively adapt to the variability in the flow direction. 

%%%% from sample input file:
%%%\vskip6pt
%%%\disclaimer{Insert disclaimer text here.}
%%%\ethics{Insert ethics text here.}
%%%\dataccess{Insert data access text here.}
%%%\ack{Insert acknowledgment text here.}
%%%\funding{Insert funding text here.}
%%%\conflict{Insert conflict text here.}

\vskip6pt
%\dataccess{All relevant information and data for reproducing and verifying the results of this work are contained in the manuscript.}
%\conflict{We declare we have no competing interest.}
%\contrib{S.M. and E.L. designed the research and G.O.A. carried out the numerical simulations and analysed the data. All authors drafted the manuscript.}
%\ack{Non applicable (sponsor acknowledged in the dedicated section).}
%\funding{The authors gratefully acknowledge  the support of Electricit\'e de France (EDF) for their support through the `Chaire Energies Durables' at Ecole Polytechnique.}
%\ethics{Non applicable (numerical work only).}

\appendix
\numberwithin{equation}{section}

\section{Local frame definition}\label{sec:axes}
The local orthonormal basis in Fig.~\ref{fig:geometry_3d_cable} is given by
\begin{align}
\begin{split}
\left[
\begin{array}{c}
\myvec{e}_r \\
\myvec{e}_\theta \\
\myvec{e}_\varphi
\end{array}
\right]
&=
\left[
\begin{array}{ccccc}
 \cos\theta\,\cos\varphi &\,&        \sin\varphi       &\,&  -\sin\theta\,\cos\varphi    \\
       -\sin\theta       &\,&             0            &\,&         -\cos\theta          \\
-\cos\theta\,\sin\varphi &\,&        \cos\varphi       &\,&   \sin\theta\,\sin\varphi 
\end{array}
\right]
\cdot
\left[
\begin{array}{c}
\myvec{e}_X \\
\myvec{e}_Y \\
\myvec{e}_Z
\end{array}
\right]
\end{split}
\label{eq:static_def_loc_frame}
\end{align}

\section{Straight Cable: Direct and Adjoint Equations}

\label{sec:adjoint}
For a straight cable with distributed harvester, the only non-zero component of the displacement is $z$, and $z$ and $q$ are governed by
\begin{align}
\ddot{z} + \left( \sigma + \frac{\gamma}{\mu} \right) \, \dot{z}-\frac{1}{\pi^2 u^2} \, z'' &= M \, q
\label{eq:VIV_straight_distrib_dless_EOM_cable}
\\
\ddot{q} + \epsilon \, \left(q^2-1\right) \, \dot{q}+q &= A \, \ddot{z}
\label{eq:VIV_straight_distrib_dless_EOM_wake}
\end{align}
and the boundary and initial conditions are $\left.z\right|_{t=0}=\left.\dot{z}\right|_{t=0}=\left.\dot{q}\right|_{t=0}=0$, $\left.q\right|_{t=0}=q_0(x)$ and $\left.z\right|_{x=0}=\left.z\right|_{x=1}=0$.

We choose a final time $t_f$ that is much larger than the saturation time and long enough to ensure that the time-average operator involved in the definition of efficiency is converged:  
\begin{align}
\eta = 16\,\mu\,\pi^3\,S_T^3\,\int_{0}^{1}\sigma(x)\,\left[\frac{1}{t_f}\int_{t=0}^{t_f}\dot{z}^2\,dt\right]\,dx
\label{eq:VIV_straight_distrib_efficiency}
\end{align}
The definition of $\eta$ in Eq.~\eqref{eq:VIV_straight_distrib_efficiency} is used here for its convenience to derive adjoint equations.

Using Eq.~\eqref{eq:VIV_straight_distrib_efficiency}, the gradient of $\eta$ with respect to the function $\sigma$ is the function of $x$:
\begin{align}
\nabla_{\sigma}[\eta](x) &= 16\,\mu\,\pi^3\,S_T^3\,\frac{1}{t_f}\int_{t=0}^{t_f}\dot{z}\,\left(\dot{z}-z^\dagger\right)\,dt
\label{eq:VIV_straight_dless_EOM_wake}
\end{align}
\noindent where the variables $z^\dagger$ and $q^\dagger$ satisfy the adjoint equations
\begin{align}
\ddot{z}^\dagger - \left( \sigma + \frac{\gamma}{\mu} \right) \, \dot{z}^\dagger-\frac{1}{\pi^2 u^2} \, \left(z^\dagger\right)'' &= A \, \ddot{q}^\dagger-2\,\sigma\,\ddot{z}
\label{eq:VIV_straight_distrib_dless_adjoint_cable}
\\
\ddot{q}^\dagger - \epsilon \, \left(q^2-1\right) \, \dot{q}^\dagger + q^\dagger &= M \, z^\dagger
\label{eq:VIV_straight_distrib_dless_adjoint_wake}
\end{align}
\noindent with the final and boundary conditions $z^\dagger|_{t=t_f}=\dot{q}^\dagger|_{t=t_f}=q^\dagger|_{t=t_f}=0$, $\dot{z}^\dagger|_{t=t_f}=-2\,\sigma\,\dot{z}|_{t=t_f}$, $z^\dagger|_{x=0}=z^\dagger|_{x=1}=0$. Note that the functions $z$ and $q$ appearing in Eqs.~\eqref{eq:VIV_straight_distrib_dless_adjoint_cable} and \eqref{eq:VIV_straight_distrib_dless_adjoint_wake} and in the final condition for $z^\dagger$ are the solutions to the direct problem Eqs.~\eqref{eq:VIV_straight_distrib_dless_EOM_cable} and \eqref{eq:VIV_straight_distrib_dless_EOM_wake}.

\section{Rigid Cylinder}
For a rigid cylinder, the governing equations for the cross-flow (vertical) displacement $z$ and wake variable $q$ are
\begin{align}
\ddot{z} + \left( \sigma + \frac{\gamma}{\mu} \right) \, \dot{z} +\frac{1}{u^2} \, z &= M \, q
\label{eq:VIV_rigid_dless_struct}
\\
\ddot{q} + \epsilon \, \left(q^2-1\right) \, \dot{q} + q &= A \, \ddot{z}
\label{eq:VIV_rigid_dless_wake}
\end{align}
\noindent where $\gamma$, $\mu$, $M$ have the same definition as for the flexible structure (Eqs.~\eqref{eq:gamma_mu_def} and \eqref{eq:VIV_3d_dless_lift}) and $u$ and $\sigma$ are defined consistently with the flexible case as $u=f/f_0$ and $\sigma=R/(2\pi f m_T)$.

\end{document}